\documentstyle[epsfig,amsmath,amssymb]{mn}

\newcommand{\df}{{\rm d}}
\newcommand{\e}{{\rm e}}
\newcommand{\half}{\tfrac{1}{2}}
\newcommand{\gccm}{\mathrm{\,g\,cm}^{-3}}
\newcommand{\comments}[1]{\bigskip\parbox[t]{0.9\linewidth}{\small{#1}}}
\newcommand{\HVpn}{HV$_{\rm pn}$}
\newcommand{\RHFpn}{RHF$_{\rm pn}$}
\newcommand{\glen}{G$_{300}$}
\newcommand{\glenpi}{G$^\pi_{300}$}
\newcommand{\UVU}{UV$_{14}$+UVII}
\newcommand{\UVT}{UV$_{14}$+TNI}
\newcommand{\KB}{G$^{\rm K240}_{\rm B180}$}
\newcommand{\KM}{G$^{\rm K240}_{\rm M78}$}
\newcommand{\KMu}{G$^{\rm K240}_{\rm M78u}$}
\newcommand{\Kpn}{G$^{\rm K240}_{\rm pn}$}
\newcommand{\SMA}{SM$_{\rm B145}$}
\newcommand{\SMB}{SM$_{\rm B160}$}

\begin{document}

\title[Quasi Periodic Oscillations and Constraints on the Equation of
State]{Quasi Periodic Oscillations in Low Mass X-Ray Binaries and
Constraints on the Equation of State of Neutron Star Matter}

\author[Ch. Schaab and M.\,K. Weigel]{Christoph Schaab and Manfred K. Weigel \\
  Institut f{\"u}r theoretische Physik,
  Ludwig-Maximilians Universit{\"a}t M{\"u}nchen, Theresienstr. 37,
  D-80333 M{\"u}nchen, Germany \\
  email: schaab@gsm.sue.physik.uni-muenchen.de}

\maketitle
\begin{abstract}
Recently discovered quasi periodic oscillations in the X-ray
brightness of low mass X-ray binaries are used to derive constraints
on the mass of the neutron star component and the equation of state of
neutron star matter. The observations are compared with models of
rapidly rotating neutron stars which are calculated by means of an
exact numerical method in full relativity. For the equations of state
we select a broad collection of models representing different
assumptions about the many-body structure and the complexity of the
composition of super dense matter. The mass constraints differ from
their values in the approximate treatment by $\sim 10$~\%. Under the
assumption that the maximum frequency of the quasi periodic
oscillations originates from the innermost stable orbit the mass of the
neutron star is in the range: $M\sim 1.92-2.25\,M_\odot$. Especially
the quasi periodic oscillation in the Atoll-source 4U\,1820-30 is only
consistent with equations of state which are rather stiff at high
densities which is explainable, so far, only with pure
nucleonic/leptonic composition. This interpretation contradicts the
hypothesis that the protoneutron star formed in SN 1987A collapsed to
a black hole, since this would demand a maximum neutron star mass
below $1.6\,M_\odot$. The recently suggested identification of quasi
periodic oscillations with frequencies around 10~Hz with the
Lense-Thirring precession of the accretion disk is found to be
inconsistent with the models studied in this work, unless it is
assumed that the first overtone of the precession is observed.
\end{abstract}


\section{Introduction}

Neutron stars contain matter in one of the densest forms found in the
universe. Their central density ranges from a few time the density of
normal nuclear matter to about one order of magnitude higher,
depending on the star's mass and the equation of state (EOS). Neutron
stars therefore provide us with a powerful tool for exploring the
properties of such dense matter. In the last decades, this tool was
applied to, among other topics, the determination of the EOS of dense,
charge neutral, $\beta$-equilibrated matter by means of comparing the
theoretical predicted properties with observations of neutron
stars. This was attempted by studying, for example, the maximum stable
star mass \cite{Kerkwijk95a}, the minimum rotation period
\cite{Friedman86a,Weber92a}, or the thermal behaviour (Tsuruta 1966
\nocite{Tsuruta66}, Schaab et al. 1996 \nocite{Schaab95a}, Page 1997
\nocite{Page97a}; see Balberg, Lichtenstadt \& Cook 1998
\nocite{Balberg98a} for a recent review).

Recently, Strohmayer et al. \shortcite{Strohmayer96a} and Van der Klis
et al. \shortcite{VanDerKlis96a} discovered with the Rossi X-ray
Timing Explorer (RXTE) kilohertz quasi-periodic oscillations (QPOs) in
the X-ray brightness of low-mass X-ray binaries (LMXBs, see van der
Klis 1997 \nocite{VanDerKlis97a} for a recent review). In subsequent
observations, three QPOs were often detected simultaneously in a given
source. The frequency separation between both QPOs is almost constant,
although the frequencies of the two QPOs themselves vary by several
hundred Hertz. Up to now, the only exceptions are the Atoll sources
4U\,1608-52 and 4U\,1735-44 and the Z-source Scorpius X-1 in which the
frequency separation varies with the luminosity by roughly $\pm
15$\,\% \cite{Mendez98b,VanDerKlis97b,Ford98c}. Psaltis et
al. \shortcite{Psaltis98a} found that the QPO data of the other LMXBs
are consistent both with being constant and with varying by a similar
fraction as in the two sources quoted before. The frequency of
oscillations during type I X-ray bursts detected in some sources are
consistent with the frequency separation of the two QPOs or with its
first overtone. Only in the source 4U\,1636-536 the averaged frequency
separation, $\Delta\nu=251$~Hz, and the half of the frequency in type
I bursts, $\nu_{\rm burst}=581$~Hz, differ by approximately 15\,\%
\cite{Mendez98c}. In the power spectrum of bursts of the same source
Miller \shortcite{Miller98c} found a second signal at 290~Hz~$\sim 1/2
\nu_{\rm burst}$. These observations strongly favour the beat
frequency model where the frequency separation between the two QPOs
originates from the stellar spin, whereas the higher QPO is produced by
accreting gas in a stable, nearly circular orbit around the neutron
star. Though, it has to be clarified how the slightly varying
frequency separation and the small deviation of the frequency
separation from the burst frequency can be incorporated into this
model.

Beside the two kilohertz QPOs and the burst oscillations, also QPOs
with frequencies of a few tens of Hertz were detected in some
sources. Their frequencies correlate with the high frequency kilohertz
QPOs. These low frequency QPOs were interpreted by Stella \& Vietri
\shortcite{Stella97a} to originate from the Lense-Thirring precession of the
accreting disc due to the frame dragging effect of the rapidly
spinning neutron star \cite{Lense18a}.

Both, the identification of the high frequency kilohertz QPOs with the
orbital frequency of a stable circular orbit and the low frequency
QPOs with the frame dragging frequency of the same orbit allow to
constrain the mass of the neutron star and also the EOS of neutron
star matter \cite{Lamb97a,Stella97a}. If the evidence for the
detection of QPOs of the innermost stable circular orbit in the
sources 4U\,1608-52, 4U\,1636-536 \cite{Kaaret97a}, and 4U\,1820-30
\cite{Zhang98c} can be confirmed by future observations, the
constraints are rather severe, allowing only a few stiff EOSs.

The frequency separation of the two kilohertz QPOs show that the
neutron star in LMXBs are rapidly rotating with periods ranging from
2.5~ms to 4~ms. It can therefore be expected that the geometry of the
neutron star and its exterior space time is non-spherical. Since the
innermost stable orbit is located at only a few kilometres above the
star's surface, the deviation from the Kerr space time are 
large and should not be neglected. In order to compare theoretical
neutron star models with QPO observations, a completely general
relativistic calculation of the rotating neutron star structure and
space time geometry is therefore necessary.

In order to study the impact and the discrimination power of the QPO
data in greater detail, we select a broad collection of modern EOSs,
which were obtained utilising numerous assumptions about the dynamics
and composition of super dense matter.  To mention several, these are:
the many-body technique used to determine the EOS; the
model for the nucleon-nucleon interaction; description of electrically
charge neutral neutron star matter in terms of either only neutrons
and protons in generalised chemical equilibrium ($\beta$ equilibrium)
with electrons and muons, or nucleons, hyperons and more massive
baryon states in $\beta$ equilibrium with leptons; hyperon coupling
strengths in matter; inclusion of meson ($\pi$, $K$) condensation;
treatment of the transition of confined hadronic matter into quark
matter; and assumptions about the true ground state of strongly
interacting matter (i.e., absolute stability of strange quark matter
relative to baryon matter).

The paper is organised as follows. In Sect. \ref{sec:equations} we
summarise the equations which govern the space time structure and
compare the approximate values of the orbital frequencies, the radius
of the innermost stable orbit, and the Lense-Thirring precession
frequencies with the respective values from the exact numerical
solution of Einstein's equations. The physics of the EOSs is discussed
in Sect. \ref{sec:eos}. The high frequency kilohertz QPOs and their
interpretation in combination with their compatibility with the
different EOSs are discussed in Sect. \ref{sec:constr1}. The
implications of the identification of the low frequency QPOs with
Lense-Thirring precession are presented in Sect. \ref{sec:constr2}. We
summarise our results, the constraints to the neutron star masses, and
the conclusions concerning the neutron star EOS in
Sect. \ref{sec:concl}.

\section{Space Time Around Rapidly Rotating Neutron Stars} 
\label{sec:equations}

\subsection{Einstein Equations} 

The stationary, axis-symmetric, and asymptotic flat metric in quasi
isotropic coordinates reads
\begin{equation} \label{eq:cool.metric}
  \df s^2 = -\e^{2\nu}\df t^2+\e^{2\phi}(\df\varphi-N^\varphi \df t)^2 
  +\e^{2\omega}(\df r^2+r^2\df\theta^2),
\end{equation}
where the metric coefficients $g_{\mu\nu}=g_{\mu\nu}(r,\theta)$ are
functions of $r$ and $\theta$ only. The metric coefficients are
determined by the Einstein equation ($c=G=1$)
\begin{equation}
  {\bf G} = 8\pi{\bf T},
\end{equation}
and the energy-momentum conservation
\begin{equation}
  {\bf\nabla}\cdot{\bf T}=0,
\end{equation}
where ${\bf T}= (e+p){\bf u}\otimes{\bf u}+p\,{\bf g}$ is the
stress-energy tensor of an ideal fluid with the 4-velocity
\begin{equation} \label{eq:def.u}
  {\bf u} = \Gamma\e^{-\nu}{\bf e}_t + \Omega\Gamma\e^{-\nu}{\bf e}_\varphi,
\end{equation}
the energy density $e$, and the pressure $p$.  The Lorentz factor
$\Gamma$ is given by ${\bf u}\cdot{\bf u}=-1$, hence
\begin{equation} \label{eq:gamma}
  \Gamma =
  \left(1-\e^{2(\phi-\nu)}\left(\Omega-N^\varphi\right)^2\right)^{-\half}.
\end{equation}
$\Omega=u^\varphi / u^t$ is the angular velocity of the fluid with
respect to an observer at infinity. The proper velocity $U$ of the
fluid with respect to the local Eulerian Observer $\mathfrak{O}_0$
\cite{Smarr78a} is given by the equation
\begin{equation} \label{eq:proper.vel}
  U = \frac{1}{g_{\varphi\varphi}^{1/2}\Gamma}{\bf e}_\varphi\cdot{\bf u}
    = \e^{\phi-\nu}\left(\Omega-N^\varphi\right).
\end{equation}
Note that if the fluid were at rest with respect to the Observer
$\mathfrak{O}_0$, $U=0$, it would not necessarily be at rest for an inertial
observer at infinity, since $\Omega=N^\varphi\neq 0$. This phenomena
of \emph{dragging of the inertial frame} was first studied by Lense
\& Thirring \shortcite{Lense18a} and turns out to be important for the
investigation of the Lense-Thirring precession of the accreting disc
(see Sect. \ref{sec:constr2}).

The four non-trivial Einstein equations together with the
energy-momentum conservation are solved via a finite difference scheme
\cite{Schaab97a}. We follow Bonazzola et al. \shortcite{Bonazzola93a} in compactifying
the outer space to a finite region by using the transformation $r
\rightarrow 1/r$. The boundary condition of approximate flatness can
then be exactly fulfilled. The neutron star model is uniquely
determined by fixing one of the parameters: central density,
gravitational mass, baryon number, Kepler orbiting frequency at the
star's equator or at the marginally stable radius, as well as one of
the parameters: angular velocity, angular momentum, or stability
parameter $t=T/|W|$. The models of maximum mass and/or maximum
rotation velocity can also be calculated.

\subsection{Stable Circular Orbits} \label{ssec:stable}

Since ${\bf e}_t$ and ${\bf e}_\varphi$ are Killing vectors, the
components $p_t$ and $p_\varphi$ of the 4-momentum of a freely falling
particle are constants of motion and can be identified with the
negative of the specific energy $E$ (in units of the rest mass $m_0$)
and the specific angular momentum $L$, respectively. For a particle
motion confined to the equatorial plane the geodesic equation
$p_\mu^\mu=-1$ yields:
\begin{equation}
  \e^{-2\omega}p_r^2 = -1+\e^{-2\nu}(E-N^\varphi L)^2-\e^{-2\phi}L^2 
        = \e^{-2\nu}V(E,L,r),
\end{equation}
where $V$ is the effective potential for the particle motion
\cite[pp. 655]{Misner73}. A circular orbit, $p_r=0$, is given by the
expression \eqref{eq:def.u} for the 4-velocity of the fluid inside the
star. Then, the specific energy $E$ and angular momentum $L$ can be
expressed in terms of the Lorentz factor $\Gamma$ (Eq. \ref{eq:gamma})
and the proper velocity $U$ (Eq. \ref{eq:proper.vel}):
\begin{align}
  E &= \Gamma\left(\e^\nu+\e^\phi N^\varphi U\right) \\
  L &= \e^\phi\Gamma U.
\end{align}
The circular orbit exists if $V_{,r}=0$, thus\footnote{We use the usual
convention, that $\cdot_{,\mu}$ represents the partial differential
$\partial/ \partial x^\mu$.}
\begin{equation}
  -\phi_{,r}U^2+\e^{\phi-\nu}N^\varphi_{,r}U+\nu_{,r} = 0.
\end{equation}
The proper velocity $U$ of a particle corotating with the star is given by
\begin{equation}
  U = \frac{\e^{\phi-\nu}N^\varphi_{,r}
        +\sqrt{\e^{2(\phi-\nu)}(N^\varphi_{,r})^2+4\phi_{,r}\nu_{,r}}}
        {2\phi_{,r}}.
\end{equation}
The circular orbit is stable, if $V_{,rr}<0$ with (see also Cook,
Shapiro \& Teukolksky 1992 \nocite{Cook92a} and Datta, Thampan \&
Bombacci 1998 \nocite{Datta98a})
\begin{multline}
  V_{,rr} = 2\e^{2\nu}\Gamma^2\left(\e^{\phi-\nu}UN^\varphi_{,rr}
        -\e^{2(\phi-\nu)}U^2(N^\varphi_{,r})^2+\nu_{,rr} \right. \\
        \left. +2(\nu_{,r})^2
        -U^2\phi_{,rr}+2U^2(\phi_{,r})^2-4\nu_{,r}\phi_{,r}U^2\right).
\end{multline}
The zero of $V_{,rr}$ determines the radius $r^{\rm ms}$ of the
innermost stable or marginally stable circular orbit. The Kepler
frequency $\nu_{\rm K}$ of a circular orbit as measured by a distant
observer is given by the proper velocity $U$ through the expression
\begin{equation}
  \nu_{\rm K} = \frac{1}{2\pi}\left(\e^{\nu-\phi}U+N^\varphi\right),
\end{equation}
and the Lense-Thirring precession frequency by
\begin{equation}
  \nu_{\rm LT} = \frac{1}{2\pi}N^\varphi.
\end{equation}
The respective values at the innermost stable orbit with radius
$r^{\rm ms}$ are denoted by $\nu_{\rm K}^{\rm ms}$ and $\nu_{\rm
LT}^{\rm ms}$, respectively. The circumferential radius $r_{\rm circ}$
of the object is linked to the equatorial coordinate radius $r$ by the
expression
\begin{equation}
  r_{\rm circ} = \e^\phi r.
\end{equation}

\subsection{Approximate Expressions}

Though we shall use the exact expressions for $\nu_{\rm K}$, $\nu_{\rm
LT}$, $\nu_{\rm K}^{\rm ms}$, and $\nu_{\rm LT}^{\rm ms}$ in our
investigation, we give also, for the purpose of comparison, the
approximate expressions valid to first order of the dimensionless
angular momentum $j=J/M^2$, where $J$ and $M$ denotes the angular
momentum and gravitational mass, respectively. The approximate
expressions have the advantage to constrain the mass and the angular
momentum independently of the EOS. To first order in $j$ the
corresponding expression are \cite{Miller98a}:
\begin{align}
  \nu_{\rm k} &= \frac{1}{2\pi}\left(\frac{M}{r_{\rm circ}^3}\right)^{\half}
        \left(1-\left(\frac{M}{r_{\rm circ}}\right)^{\tfrac{3}{2}}j\right), \\
  \nu_{\rm LT} &= \frac{2}{2\pi}\frac{M^2}{r_{\rm circ}^3}j, \\
  r_{\rm circ}^{\rm ms} &= 
        6M\left(1-\left(\tfrac{2}{3}\right)^{\tfrac{3}{2}}j\right), \\
  \nu_{\rm K}^{\rm ms} &= \frac{6^{-\tfrac{3}{2}}}{2\pi}\frac{1}{M}
        \left(1+11\times 6^{-\tfrac{3}{2}}j\right) \label{eq:appr.Kms}, \\
  \nu_{\rm LT}^{\rm ms} &= \frac{6^{-3}}{\pi}\frac{j}{M}.
\end{align}

\section{Equations of State} \label{sec:eos}

\subsection{Neutron Star Matter}

The EOS of neutron star matter is the basic input quantity to the
structure equations discussed in Sect. \ref{sec:equations}. Its
knowledge over a wide range of densities is necessary. Whereas the
density at the star's surface corresponds to the density of iron, the
density in the centre of a very massive star can reach 15 times the
density of normal nuclear matter. Since neutron star matter in
chemical equilibrium (i.e.\ generalized $\beta$-equilibrium) is highly
isospin-asymmetric and may carry net-strangeness its properties cannot
be explored in laboratory experiments. Therefore, one is left with
models for the EOS which depend on theoretically motivated assumptions
and/or speculations about the behaviour of super dense matter. One
main source of uncertainty is the competition between non-relativistic
versus relativistic models. Though both treatments give satisfactory
results for normal nuclear densities, they provide quite different
results if one extrapolates to higher densities (see, e.g, Huber et
al. 1998\nocite{Huber97a}). Moreover, one encounters strong
differences in the high density regime depending on the underlying
dynamics, the many-body approximation and the assumptions about the
composition. The simplest approach describes neutron star matter by
pure neutron matter. Since pure neutron matter is certainly not the
true ground state of neutron star matter, it will quickly decay by
means of the weak interaction into chemically equilibrated neutron
star matter, whose fundamental constituents -- besides neutrons -- are
protons, hyperons and possibly more massive baryons. Even a transition
to quark matter (so-called hybrid stars) and the occurrence of pion-
or kaon-condensates is possible.

The cross section of a neutron star can be split roughly into three
distinct regimes \cite{Boerner73b,Weber91}. The first one is the
star's outer crust, which consists of a lattice of atomic nuclei and a
Fermi liquid of relativistic, degenerate electrons. The inner crust
extends from neutron drip density, $\rho=4.3\times 10^{11}\gccm$, to a
transition density of about $\rho_{\rm tr}=1.7\times 10^{14}\gccm$
\cite{Pethick95}.  Beyond this transition density $\rho_{\rm tr}$
one enters the star's third regime, that is, its core where all atomic
nuclei have dissolved into their constituents, protons and
neutrons. Furthermore, as outlined just above, due to the high Fermi
pressure the core will contain hyperons, eventually more massive
baryon resonances, and possibly a gas of free up, down and strange
quarks.

The EOS of the outer and inner crust has been studied in
several investigations and is rather well known.  We shall adopt for
these regions the models derived by Haensel \& Pichon
\shortcite{Haensel94a} and by Negele \& Vautherin \shortcite{Negele73},
respectively.  The models for the EOS of the star's core
are discussed in detail in Schaab et al. \shortcite{Schaab95a}.

An overview of the collection of EOSs used in this paper
for the core region is given in Tab. \ref{tab:eos}. We have chosen a
representative collection of different models in order to check which
EOSs are compatible with the QPOs and their theoretical
interpretation.
\begin{table*} 
  \centering \caption{Dynamics and approximation schemes for EOSs
  derived for the cores of neutron stars \label{tab:eos}} \smallskip
  \scriptsize
\begin{tabular}{ccccc}
  \hline
  \hline
  EOS & Composition & Interaction & Many body approach & Reference \\
  \hline
  UV$_{14}$+UVII 
  & p, n, e$^-$, $\mu^-$ 
  & Urbana V$_{14}$ and Urbana VII 
  & NRV
  & Wiringa, Fiks \& Fabrocini \shortcite{Wiringa88} \\
  UV$_{14}$+TNI 
  & p, n, e$^-$, $\mu^-$ 
  & Urbana V$_{14}$ and TNI
  & NRV
  & Wiringa et al. \shortcite{Wiringa88} \\
  TF
  & p, n, e$^-$, $\mu^-$ 
  & TF96
  & Thomas-Fermi model
  & Strobel et al. \shortcite{Strobel96a} \\
  \HVpn
  & p, n, e$^-$, $\mu^-$
  & exchange of $\sigma$, $\omega$, $\rho$ 
  & RH 
  & Weber \& Weigel \shortcite{Weber89} \\
  HV
  & p, n, $\Lambda$, $\Sigma^{\pm,0}$, $\Xi^{0,-}$, e$^-$, $\mu^-$
  & exchange of $\sigma$, $\omega$, $\rho$ 
  & RH 
  & Weber \& Weigel \shortcite{Weber89} \\
  RH1
  & p, n, $\Lambda$, $\Sigma^{\pm,0}$, $\Xi^{0,-}$, $\Delta^{-,0,+,++}$, e$^-$, $\mu^-$
  & exchange of $\sigma$, $\omega$, $\pi$, $\rho$
  & RBHF+RH 
  & Huber et al. \shortcite{Huber97a} \\
  \RHFpn
  & p, n, e$^-$, $\mu^-$
  & exchange of $\sigma$, $\omega$, $\pi$, $\rho$
  & RBHF+RH 
  & Huber et al. \shortcite{Huber97a} \\
  RHF1
  & p, n, $\Lambda$, $\Sigma^{\pm,0}$, $\Xi^{0,-}$, $\Delta^{-,0,+,++}$, e$^-$, $\mu^-$
  & exchange of $\sigma$, $\omega$, $\pi$, $\rho$
  & RBHF+RHF 
  & Huber et al. \shortcite{Huber97a} \\
  RHF8
  & p, n, $\Lambda$, $\Sigma^{\pm,0}$, $\Xi^{0,-}$, $\Delta^{-,0,+,++}$, e$^-$, $\mu^-$
  & exchange of $\sigma$, $\omega$, $\pi$, $\rho$
  & RBHF+RHF 
  & Huber et al. \shortcite{Huber97a} \\
  \glen
  & p, n, $\Lambda$, $\Sigma^{\pm,0}$, $\Xi^{0,-}$, e$^-$, $\mu^-$
  & exchange of $\sigma$, $\omega$, $\rho$
  & RH 
  & Glendenning \shortcite{Glendenning89} \\
  \glenpi
  & p, n, $\Lambda$, $\Sigma^{\pm,0}$, $\Xi^{0,-}$, e$^-$, $\mu^-$, $\pi$-condensation
  & exchange of $\sigma$, $\omega$, $\rho$
  & RH 
  & Glendenning \shortcite{Glendenning89} \\
  \Kpn   
  & p, n, e$^-$, $\mu^-$
  & exchange of $\sigma$, $\omega$, $\rho$
  & RH
  & Glendenning \shortcite{Glendenning95b} \\
  \KM
  & p, n, $\Lambda$, $\Sigma^{\pm,0}$, $\Xi^{0,-}$, e$^-$, $\mu^-$
  & exchange of $\sigma$, $\omega$, $\rho$
  & RH
  & Glendenning \shortcite{Glendenning97c} \\
  \KMu
  & p, n, $\Lambda$, $\Sigma^{\pm,0}$, $\Xi^{0,-}$, e$^-$, $\mu^-$
  & exchange of $\sigma$, $\omega$, $\rho$
  & RH
  & Glendenning \shortcite{Glendenning95b} \\
  \KB   & p, n, $\Lambda$, $\Sigma^{\pm,0}$, $\Xi^{0,-}$, e$^-$, $\mu^-$, quarks
  & exchange of $\sigma$, $\omega$, $\rho$
  & RH
  & Glendenning \shortcite{Glendenning97c} \\
  TM1-m1
  & p, n, $\Lambda$, $\Sigma^{\pm,0}$, $\Xi^{0,-}$, e$^-$, $\mu^-$
  & exchange of $\sigma$, $\omega$, $\rho$, $\phi$
  & RH 
  & Schaffner \& Mishustin \shortcite{Schaffner95a} \\
  TM1-m2
  & p, n, $\Lambda$, $\Sigma^{\pm,0}$, $\Xi^{0,-}$, e$^-$, $\mu^-$
  & exchange of $\sigma$, $\omega$, $\rho$, $\sigma^*$, $\phi$
  & RH 
  & Schaffner \& Mishustin \shortcite{Schaffner95a} \\
  NLSH1
  & p, n, $\Lambda$, $\Sigma^{\pm,0}$, $\Xi^{0,-}$, e$^-$, $\mu^-$
  & exchange of $\sigma$, $\omega$, $\rho$, $\phi$
  & RH 
  & Schaffner \& Mishustin \shortcite{Schaffner95a} \\
  NLSH2
  & p, n, $\Lambda$, $\Sigma^{\pm,0}$, $\Xi^{0,-}$, e$^-$, $\mu^-$
  & exchange of $\sigma$, $\omega$, $\rho$, $\sigma^*$, $\phi$
  & RH 
  & Schaffner \& Mishustin \shortcite{Schaffner95a} \\
  \SMA
  & u, d, s
  & Bag model
  & Fermi gas
  & Farhi \& Jaffe \shortcite{Farhi84} \\
  \SMB
  & u, d, s
  & Bag model
  & Fermi gas
  & Farhi \& Jaffe \shortcite{Farhi84} \\
  \hline\hline
\end{tabular} 
\comments{Abbreviations: NRV=non-relativistic variational method,
  RH=relativistic Hartree approximation, RHF=relativistic Hartree-Fock
  approximation, RBHF=relativistic Br\"uckner-Hartree-Fock
  approximation.}
\end{table*}

An important characteristic of relativistic models is the appearance
of a new saturation mechanism. In relativistic theories the repulsive
force is caused by the exchange of vector mesons coupled to the baryon
densities, whereas the attraction is coupled to the scalar densities
by means of the scalar mesons. Hence, the repulsion increases with
increasing density with respect to the attraction. Since
non-relativistic treatments do not distinguish both kinds of densities,
one has to introduce adhoc forces depending explicitly on density in
order to reproduce properties of saturated nuclear matter.

The potential parameters and coupling constants in the pure nucleonic
sector are adjusted to nucleon--nucleon scattering data and properties
of the deuteron in the case of the non-relativistic microscopic models
and the relativistic Br\"uckner-Hartree-Fock models. In this sense,
these models are called \emph{parameter free}. For the Hartree- and
the Hartree-Fock approximation, as well as for the Thomas-Fermi model
such an adjustment is not possible, since the free force parameters
would not yield saturation of nuclear matter. In such more
phenomenological approximations the coupling constants are adjusted to
properties of saturated nuclear matter or atomic nuclei
\cite{Weber89,Schaffner95a,Myers96a}. In a more recent and more elaborate
investigation \cite{Huber97a}, the adjustment was performed with
respect to properties of neutron star matter at 2--3 times nuclear
density known from relativistic Br\"uckner-Hartree-Fock
calculations. Relativistic Br\"uckner-Hartree-Fock calculations cannot
be performed over the whole density range at present, since an
inclusion of hyperons leads to rather complex equation systems. Even
for pure nucleonic matter, the self consistent method is numerically
stable up to 2--3 times nuclear density only \cite{Huber97a}.

Since the coupling constants of hyperons are not obtainable from
properties of normal nuclei and data of hypernuclei are relatively
scarce, one has a greater freedom in the selection of parameter sets
in the hyperonic sector. From a theoretical standpoint one may first
utilise the SU(6) symmetry of the quark model for the vector coupling
constants and adjust the $\sigma$-coupling from the binding energy of
the $\Lambda$ hyperon in nuclear matter. It turns out, however, that
this procedure gives different $\sigma$-couplings depending on the
chosen many-body approximation. Therefore, it seems reasonable to vary
both couplings with the constraint of the $\Lambda$ binding energy in
nuclear matter in such phenomenological many-body theories. A further
constraint in this procedure is the compatibility with the allowed
neutron star masses (for more details, see Huber et al. 1998
\nocite{Huber97a} and Huber 1998 \nocite{Huber98a}). The coupling
constants for the strange mesons can be obtained to a certain extent
from the data of double $\Lambda$-hypernuclei
\cite{Schaffner95a}. Batty et al. \shortcite{Batty94a} claim some doubts on
the existence of $\Sigma$ hyperons in neutron star matter. However,
the disappearance of $\Sigma$ hyperons would only slightly change the
EOS
\cite{Balberg96a}. A further open question is still the existence of
$\Delta$ isobars in the interior of neutron stars. In relativistic
Hartree treatments, the isospin unfavoured $\Delta^-$ isobar does not
appear because of the rather large $\rho$-coupling, which is necessary
to reproduce the empirical symmetry coefficient
\cite{Huber97a}. $\Delta$ isobars are therefore often neglected a
priori in relativistic Hartree treatments
\cite{Schaffner95a}. However, in relativistic Hartree-Fock
approximations, the $\rho$-coupling becomes smaller due to the
exchange contributions. For that reason, the charge favoured
$\Delta^-$ may now enter the composition depending on the behaviour of
the effective $\Delta$-mass in neutron star matter, which is quite
uncertain \cite{Huber97a}.

The possibility of a transition of confined hadronic matter into quark
matter at high densities is included in the EOS labelled
\KB~ \cite{Glendenning95b} (so-called hybrid stars). The transition
was determined for a bag constant of $B^{1/4}=180$ MeV which places
the energy per baryon of strange quark matter at 1100 MeV, well above
the energy per nucleon in saturated nuclear matter as well as in the
most stable nucleus, $^{56}{\rm Fe}$ ($E/A\approx$ 930 MeV).  This
model predicts a transition to a mixed phase consisting of quark
matter and hadronic matter at a density as low as $1.6\,n_0$. The pure
quark-matter phase is reached at a density of about $10 \, n_0$, which
is close to the central density of the heaviest and thus most compact
star constructed from such an EOS. A phase transition to pion
condensation is accounted for in \glenpi. This EOS predicts a phase
transition at $n\approx 1.3\,n_0$. The possibility of absolutely
stable strange quark matter will be considered in the following
section.

\begin{figure} \centering 
\psfig{figure=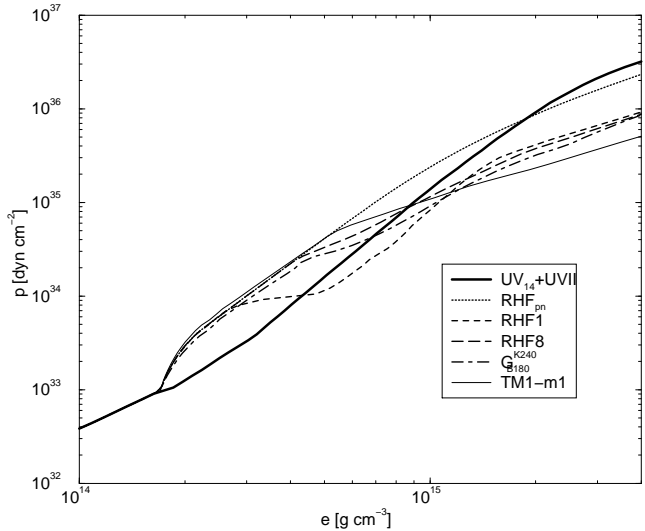,height=\linewidth,angle=-90}
\caption{Pressure-density relation for several EOSs}
\label{fig:eos}
\end{figure}
The stiffness of the EOS depends strongly on the internal degrees of
freedom. Generally, the more degrees of freedom are taken into account
the softer the EOS becomes (s. Fig. \ref{fig:eos}). A softer EOS, in
turn, leads to lower maximum neutron star masses and, for fixed mass,
to higher central densities.

\subsection{Strange Stars} \label{ssec:strangestars}

The hypothesis that strange quark matter may be the absolute ground
state of the strong interaction (not $^{56}{\rm Fe}$) has been raised
independently by Bodmer \shortcite{Bodmer71} and Witten \shortcite{Witten84}.
If the hypothesis is true, then a separate class of compact stars
could exist, which range from dense strange stars to strange dwarfs to
strange planets \cite{Weber94,Glendenning94a,Glendenning95c}. In
principle both strange and neutron stars could exist. However if
strange stars exist, the galaxy is likely to be contaminated by
strange quark nuggets which would convert via impact ``conventional''
neutron stars into strange stars
\cite{Glendenning90,Madsen91,Caldwell91}.  This would imply that the
objects known to astronomers as pulsars are probably rotating strange
matter stars and not neutron matter stars, as it is usually assumed.
Presently there is no sound scientific basis on which one can either
confirm or reject the hypothesis, so that it remains a serious
possibility of fundamental significance for compact stars
\cite{Weber96a,Weber96b,Madsen98a} plus various other physical
phenomena \cite{Aarhus91}. Below we will explore the implications of
the existence of strange stars with respect to the QPO phenomena. This
enables one to test the possible existence of strange stars and thus
draw definitive conclusions about the true ground state of strongly
interacting matter.

\begin{figure} \centering 
\psfig{figure=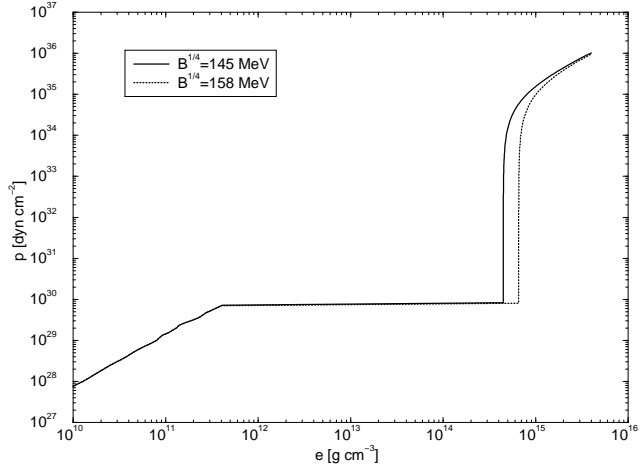,height=\linewidth,angle=-90}
\caption{Pressure-density relation for strange stars with crust. The bag
  constant is $B^{1/4}=145$~MeV or 158~MeV, respectively, the mass of
  the strange quark is $m_s=100$~MeV. }
\label{fig:sm:eos1}
\end{figure}

As pointed out by Alcock et al. \shortcite{Alcock86}, a strange star can
carry a solid nuclear crust whose density is strictly limited by
neutron drip.  This possibility is caused by the displacement of
electrons at the surface of the strange matter core, which generates an
electric dipole layer there.  It can be sufficiently strong to
produce a gap between ordinary atomic matter (crust) and the
quark-matter surface, which prevents a conversion of ordinary atomic
matter into the assumed lower-lying ground state of strange matter.
Obviously, free neutrons, being electrically charge neutral, cannot
exist in the crust, because they do not feel the Coulomb barrier and
thus would gravitate toward the strange-quark matter core, where they
are converted by hypothesis into strange matter.  Consequently, the
density at the base of the crust (inner crust density) must always be
smaller than neutron drip density.  The situation is illustrated in
Fig. \ref{fig:sm:eos1} where the EOS of a strange star
with crust is shown.

The strange-star models presented in the following are constructed for
an EOS of strange matter derived by Farhi \& Jaffe
\shortcite{Farhi84}. We shall study models with a fixed strange quark mass,
$m_{\rm s}=100$~MeV, and the two bag constants, $B^{1/4}=145$~MeV and
$B^{1/4}=158$~MeV.

\section{Kilohertz Quasi Periodic Oscillations} \label{sec:constr1}

\begin{table*} 
  \centering \caption{Observational data of kilohertz QPOs
         \label{tab:qpo.data}}
  \smallskip 
\begin{tabular}{ccccccc}
\hline\hline
  Source & Type & $\nu_{\rm QPO1}$ [Hz] & $\nu_{\rm QPO2}$ [Hz] 
  & $\Delta\nu_{\rm QPO1}$ [Hz] & $\nu_{\rm Burst}$ [Hz] 
  & References \\
  \hline
  4U\,0614+091   & Atoll         & 750--$1145\pm 10$ & 480--800  
        & $323\pm 4$            & 328\dag
        & 1 \\
  4U\,1608-52\ddag & Atoll               & 940--1125     & 650--890      
        & $233\pm 12$--$293\pm 7$&
        & 2,3,4,5,6 \\
  4U\,1636-536\ddag & Atoll              & 1147--1228    & 830--940
        & $257\pm 20$--$276\pm 10$& 581
        & 6,7,8 \\
  4U\,1728-34    & Atoll         & 988--$1058\pm 12$ & 638--716
        & $355\pm 5$            & 363   
        & 7,9 \\
  KS 1731-260   & Atoll         & 1176--$1197\pm 10$ & 900
        & $260\pm 10$           & $523.92\pm 0.05$ 
        & 10,11 \\
  4U\,1735-44    & Atoll         & 982--$1161\pm 1$ & 632--729
        & $296\pm 12$--$341\pm 7$        & 
        & 12,13 \\
  4U\,1820-30\ddag & Atoll               & 640-$1060\pm 20$ & 400--800
        & $275\pm 8$            & 
        & 14 \\
  Aql X-1       & Atoll         & 740--830      & 
        &                       & 549   
        & 15 \\
  GX 3+1        & Atoll         &               & 
        &                       &       
        & 16 \\
  4U\,1705-44    & Atoll         & $1074\pm 10$  & 776--867
        & $298\pm 11$           &       
        & 17 \\
  Sco X-1       & Z             & 872--1115     & 565--890
        & 310--230              & 
        & 18 \\
  GX 5-1        & Z             & 567--895      & 325--448
        & $298\pm 11$           &       
        & 19 \\
  GX 17+2       & Z             & 645--$1087\pm 12$ & 480--781
        & $294\pm 8$            &       
        & 20 \\
  Cyg X-2       & Z             & 731--$1007\pm 12$ & 490--780
        & $346\pm 29$           &       
        & 21 \\
  GX 349+2      & Z             & $978\pm 9$    & 712
        & $266\pm 13$           & 
        & 22 \\
  GX 340+0      & Z             & 567--$820\pm 19$ & 247--625
        & $325\pm 10$           & 
        & 23 \\
  \hline\hline
\end{tabular} \comments{\dag: QPO is statistical not significant, 
  \ddag: $\nu_{\rm QPO1}^{\rm max}$ is independent of
  luminosity. References 1: Ford et al. \shortcite{Ford97a}, 2:
  M{\'e}ndez et al. \shortcite{Mendez97a}, 3: M{\'e}ndez et
  al. \shortcite{Mendez98a}, 4: Yu et al. \shortcite{Yu97a}, 5: Yu et
  al. \shortcite{Yu97b}, 6: Kaaret et al. \shortcite{Kaaret97a}, 7:
  van der Klis \shortcite{VanDerKlis97a}, 8: Strohmayer et
  al. \shortcite{Strohmayer98b}, 9: Strohmayer et
  al. \shortcite{Strohmayer96a}, 10: Wijnands \& van der Klis
  \shortcite{Wijnands97a}, 11: Smith, Morgan \& Bradt
  \shortcite{Smith97a}, 12: Wijnands et al. \shortcite{Wijnands98b},
  13: Ford et al. \shortcite{Ford98c}, 14: Zhang et
  al. \shortcite{Zhang98c}, 15: Zhang et al. \shortcite{Zhang98a}, 16:
  Strohmayer \shortcite{Strohmayer98a}, 17: Ford, van der Klis \&
  Kaaret \shortcite{Ford98a}, 18: van der Klis \& Wijnands
  \shortcite{VanDerKlis97b}, 19: Wijnands et
  al. \shortcite{Wijnands98d}, 20: Wijnands et
  al. \shortcite{Wijnands97b}, 21: Wijnands et
  al. \shortcite{Wijnands98a}, 22: Zhang, Strohmayer \& Swank
  \shortcite{Zhang98b}, 23: Jonker et al. \shortcite{Jonker98a}}
\end{table*}
In Table \ref{tab:qpo.data} the observations of kilohertz QPOs from 16
sources are summarised. The sources can be classified into two
classes, the Atoll- and the Z-sources, depending on the form of their
colour-colour diagram. Almost all sources show simultaneously two
kilohertz QPOs. Although the frequencies of both QPOs vary by several
hundred Hertz with the accretion rate, the variation of the frequency
separation is only small.

Therefore the observations strongly favour some kind of beat-frequency
model. In such models the upper kilohertz QPO corresponds to the
Kepler rotation at the inner edge of the accretion disk. The lower QPO
corresponds to the beat frequency between the upper QPO and the spin
of the neutron star. This interpretation is supported by the fact that
the frequency separation is equal to the QPO frequency (or half of it)
in type I X-ray bursts observed in some of the Atoll sources (see
Tab. \ref{tab:qpo.data}). The question remains however where the upper
QPO is generated. Strohmayer et al. \shortcite{Strohmayer96a}
suggested that this radius is identical to the magnetospheric radius,
whereas Miller et al. \shortcite{Miller96a} propose that it is
identical to the sonic radius (see van der Klis 1997
\nocite{VanDerKlis97a} for a critical discussion of these models). In
both cases, the orbital radius has to be larger than the radius of the
innermost stable orbit (see Sect. \ref{ssec:stable}), or equivalently,
the frequency $\nu_{\rm QPO1}$ of the upper QPO has to be smaller than
the Kepler frequency at the innermost stable orbit:
\begin{equation}
  \nu_{\rm QPO1} \leq \nu_{\rm K}^{\rm ms} 
        \approx 2200 \frac{M_\odot}{M}~{\rm Hz},
\end{equation}
where the approximate expression \eqref{eq:appr.Kms} was used. The
exact expression depends on the EOS and on the spin period of the
neutron star. This inequality sets an upper limit to the mass of the
star. Obviously, the orbital radius has to be larger than the star's
radius, which is larger than the radius of the innermost stable orbit
for low star masses. This constraint sets a lower limit to the star
mass which again depends on the spin period and the EOS.

\begin{figure} \centering 
  \psfig{figure=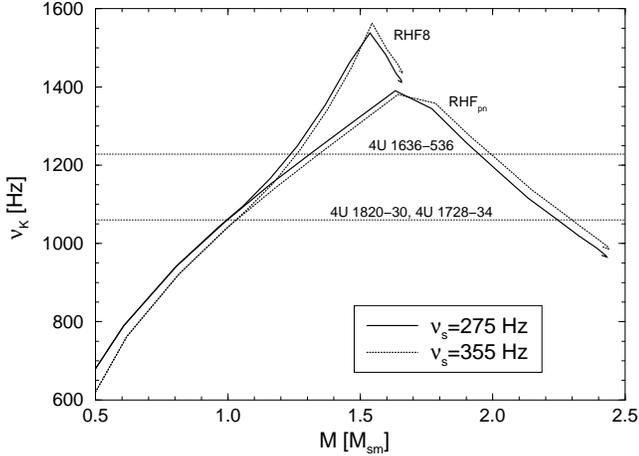,height=\linewidth,angle=-90}
  \caption{Maximally allowed Kepler frequency $\nu_{\rm K}$ as function
  of the neutron star's mass $M$ for the two EOSs \RHFpn\ and RHF8
  and two different spin frequencies $\nu_{\rm s}=275$ and 355~Hz.}
  \label{fig:qpo3}
\end{figure}
Figure \ref{fig:qpo3} shows the maximally allowed Kepler frequency as
function of the gravitational mass for two EOSs, \RHFpn\ and RHF8. If
the radius of the innermost stable orbit is smaller than the star's
radius the maximally allowed Kepler frequency is given by the Kepler
frequency at the star's equator. This kind of solution is represented
by the rising branch of $\nu_{\rm K}(M)$, since $\nu_{\rm K}$
increases with increasing gravitational mass. If the innermost stable
orbit is located outside the star the maximally allowed Kepler
frequency decreases again with increasing mass. The two EOSs differ
for $M\gtrsim 1.0\,M_\odot$ in the composition. In RHF8 hyperons are
included, whereas \RHFpn\ assumes pure nucleonic matter. RHF8 is
therefore softer than \RHFpn\ at high densities. This means also, that
RHF8 can only support smaller masses than \RHFpn. The maximally
allowed Kepler frequency depends only slightly on the neutron star's
spin frequency, as long as the spin frequency is not larger than about
ten percent of the star's Kepler frequency.  If one compares the
$\nu_{\rm K}$--curves for example with the frequency $\nu_{\rm
QPO1}^{\rm max}=1060$ of the highest kilohertz QPO observed in
4U\,1820-30 ($\nu_{\rm s}=275$~Hz), one obtains that the mass of
the neutron star has to be within the range between $M=1.0\,M_\odot$
and $2.25\,M_\odot$ (for \RHFpn) and $M_{\rm max}=1.65\,M_\odot$ (for
RHF8), respectively.

If one adopts the interpretation of the sonic-point model one expects
that the inner boundary of the accretion disc is close to the innermost
stable orbit. Indeed, the observations of the sources 4U\,1820-30, 4U
1608-52, and 4U\,1636-536 seem to confirm this interpretation, since
the QPO frequency remains constant for high mass accretion rates
\cite{Kaaret97a,Zhang98c}. Therefore, only the right intersection
point of the $\nu_{\rm K}(M)$-curve with the line $\nu_{\rm
K}=\nu_{\rm QPO1}^{\rm max}$ is allowed by the
observations\footnote{The left intersection point is unlikely, since
it is expected that the lower and the upper frequency QPO cannot be
observed at the same time if the accretion disk would extend to the
neutron star surface \cite{Zhang97a}. This would be contrary to the
simultanous observation of both QPOs in several sources.}. The mass of
the neutron star in the binary system 4U\,1820-30 can now be
determined. For
\RHFpn\, the mass is equal to $M=2.25M_\odot$ whereas RHF8 would be
excluded.

\begin{figure} \centering 
  \psfig{figure=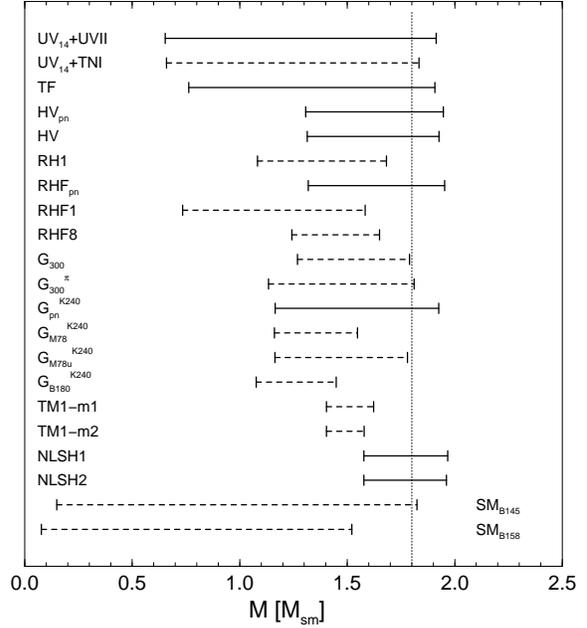,height=\linewidth} \caption{Range
  of masses for which the Kepler frequency at the equator or at the
  innermost stable orbit is larger than the highest observed QPO
  frequency $\nu_{\rm QPO1}^{\rm max}=1228$~Hz in the source 4U
  1636-536 ($\nu_{\rm s}=290$~Hz). The dashed bars refer to the
  EOSs for which models with $\nu_{\rm K}^{\rm ms}=\nu_{\rm QPO1}^{\rm
  max}$ do not exist. The approximate value $M_{\rm appr}=1.80M_\odot$
  is also shown.}  \label{fig:qpo2a}
\end{figure}
\begin{figure} \centering 
  \psfig{figure=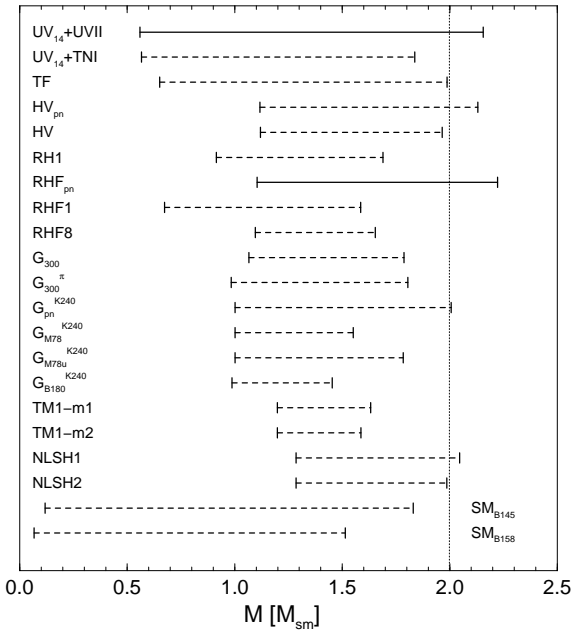,height=\linewidth} \caption{Same as
  in Fig. \ref{fig:qpo2a} but for the source 4U\,1728-34 ($\nu_{\rm
  QPO1}^{\rm max}=1100$~Hz, $\nu_{\rm s}=355$~Hz, $M_{\rm
  appr}=2.01M_\odot$).}  \label{fig:qpo2b}
\end{figure}
\begin{figure} \centering 
  \psfig{figure=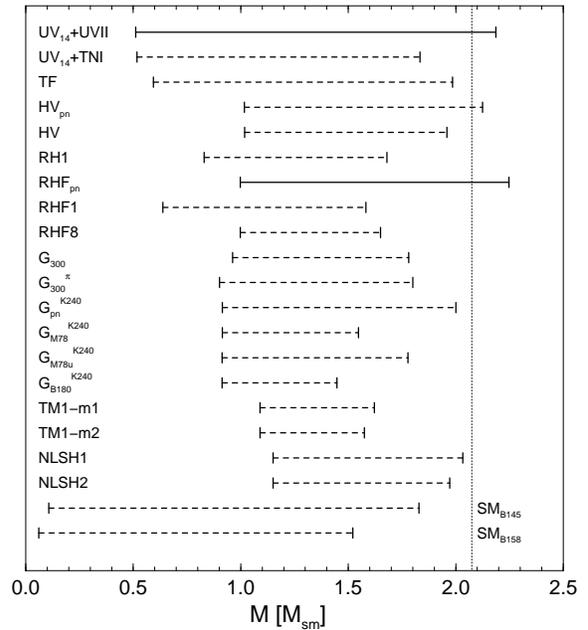,height=\linewidth} \caption{Same as
  in Fig. \ref{fig:qpo2a} but for the source 4U\,1820-30 ($\nu_{\rm
  QPO1}^{\rm max}=1060$~Hz, $\nu_{\rm s}=275$~Hz, $M_{\rm
  appr}=2.07M_\odot$).}  \label{fig:qpo2c}
\end{figure}
The figures \ref{fig:qpo2a}--\ref{fig:qpo2c} show the respective
ranges of masses for which the Kepler frequency at the equator or at
the innermost stable orbit is larger than the highest observed QPO
frequency $\nu_{\rm QPO1}^{\rm max}$ for all EOSs considered here (see
Tab. \ref{tab:eos}). The vertical line refers to the approximate mass,
which is obtained by setting $j=0$ in Eq. \eqref{eq:appr.Kms}. This
expression underestimated the upper limit of the mass. The dashed bars
refer to the EOSs for which models with $\nu_{\rm K}^{\rm ms}=\nu_{\rm
QPO1}^{\rm max}$ do not exist, i.e. whose maximally stable mass is to
small. If the interpretation of the highest observed QPO frequency in
the three sources 4U\,1820-30, 4U\,1608-52, and 4U\,1636-536 is
confirmed, the mass of the neutron star is constrained to the
respective right end of the solid bars. The EOSs, for which models
with $\nu_{\rm K}^{\rm ms}=\nu_{\rm QPO1}^{\rm max}$ do not exist
(dashed bars) are excluded.

Especially, the observation of 4U\,1820-30 agrees only with the two
nucleonic EOSs \UVU\ and \RHFpn. If the interpretation of the highest
observed QPO frequency in this source is confirmed by further
observations, the existence of hyperons, meson condensates, quark
condensates, and strange stars can be rejected. The neutron star
matter consists thus only of nucleons and leptons. At least, the
hyperon fraction (i.e. strangeness) is much smaller than predicted by
actual relativistic calculations.

\begin{figure} \centering 
  \psfig{figure=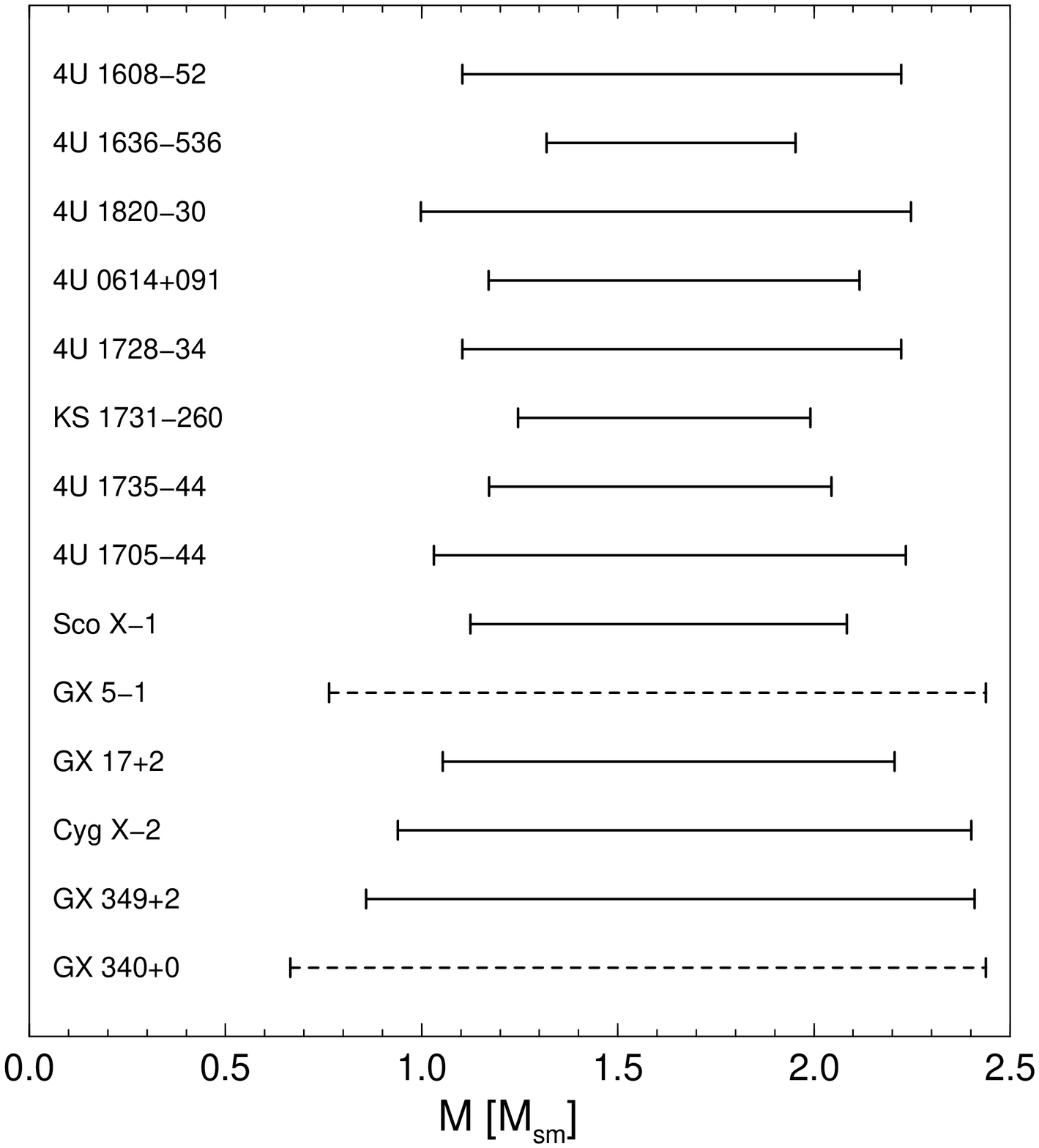,height=\linewidth} \caption{Same as
  in Fig. \ref{fig:qpo2a} but for different sources and the fixed EOS
  \RHFpn.}  \label{fig:qpo6a}
\end{figure}
\begin{figure} \centering 
  \psfig{figure=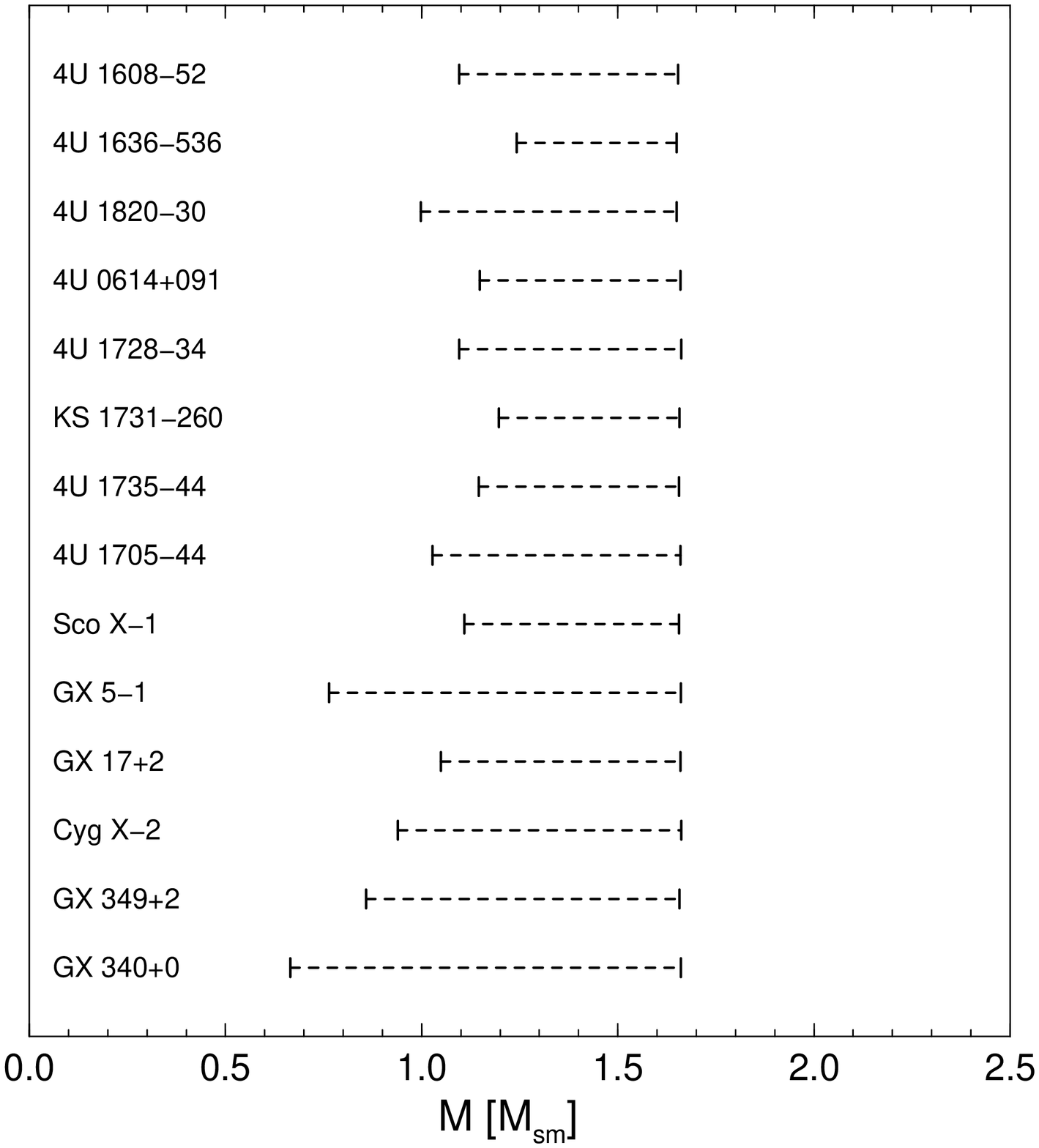,height=\linewidth} \caption{Same as
  in Fig. \ref{fig:qpo6a} but for RHF8.}  \label{fig:qpo6b}
\end{figure}
Figures \ref{fig:qpo6a} and \ref{fig:qpo6b} show the allowed ranges of
mass of the neutron stars in the respective binaries for the EOSs
\RHFpn\ and RHF8, respectively. For RHF8, the masses are limited by
the maximally stable star mass $M_{\rm max}\sim 1.65\,M_\odot$, which
depends on the spin period. All sources are consistent with a
canonical star mass of $M=1.4M_\odot$. Only if one assumes that the
highest QPO frequency is equal to the Kepler frequency at the
innermost stable orbit, the masses lie between $M=2.0\,M_\odot$ and
$2.5\,M_\odot$ (for \RHFpn).

\section{Constraints on Equation of State by Lense-Thirring Precession}
\label{sec:constr2}

\begin{figure*} \centering 
  \psfig{figure=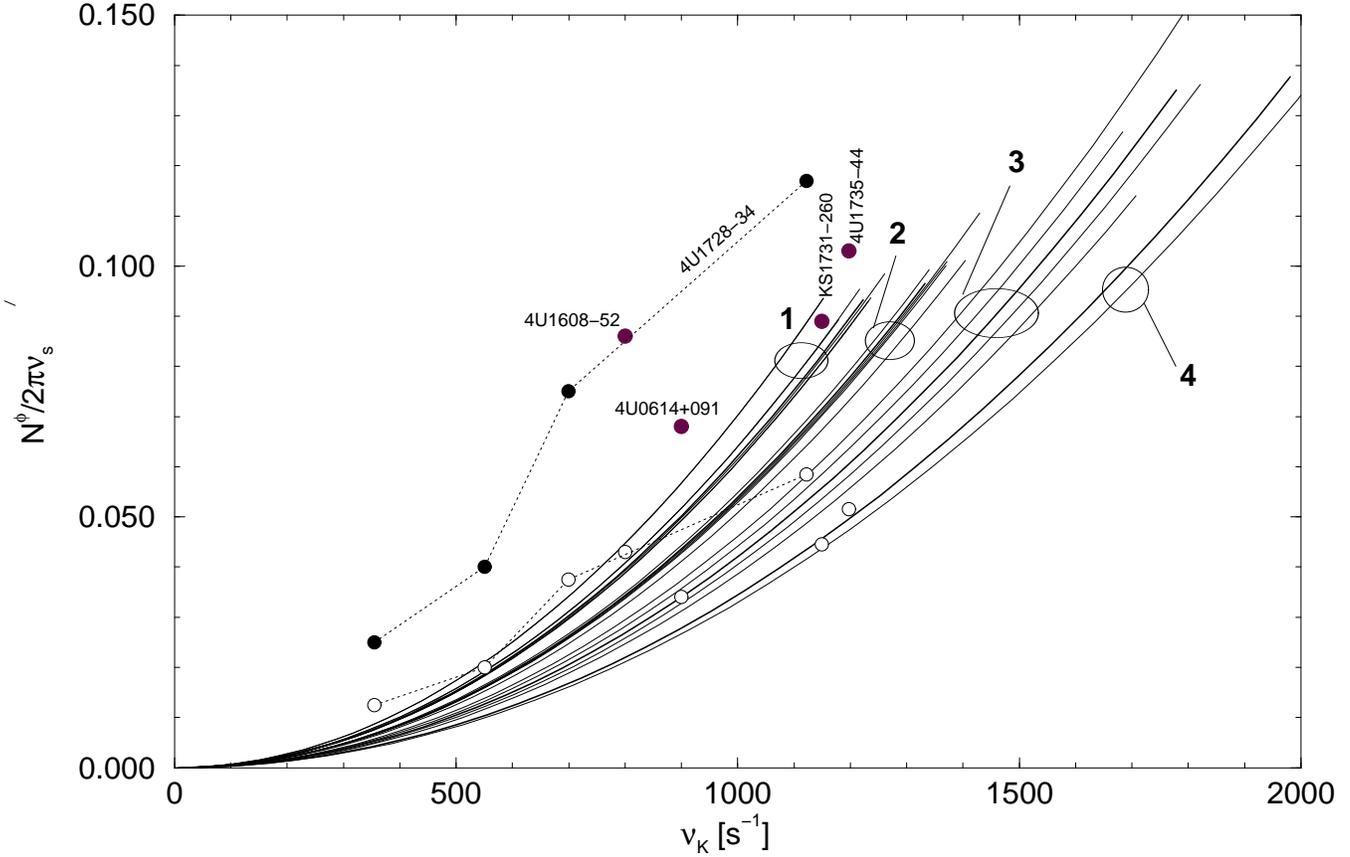,height=\linewidth,angle=-90} \caption{Ratio
  $\nu_{\rm LT}/\nu_{\rm s}$ of the Lense-Thirring precession
  frequency $\nu_{\rm LT}$ ant the spin frequency $\nu_{\rm s}\equiv
  363$~Hz as function of the Kepler frequency $\nu_{\rm K}$. The used
  mass is $M=1.4M_\odot$. The first group of curves contains the EOSs:
  NLSH1, NLSH2, TM1-m2, TM1-m1, {\bf\RHFpn}, \glen, \HVpn, and HV
  (from the left to the right; the bold typed label corresponds to the
  bold curve). Group 2: \Kpn, {\bf RHF8}, RH1, \glenpi, \KMu, and
  \KM. Group 3: \SMA, TF, {\bf \UVU}, \UVT, and \KB. Group 4: {\bf
  RHF1} and \SMB. The full (open) circles correspond to (half of)
  the observations of the Atoll-sources in Tab. \ref{tab:qpo}.}
  \label{fig:rot.qpo}
\end{figure*}

\begin{table*} 
  \centering \caption{Observational data of QPOs in the 10~Hz range. \label{tab:qpo}}
\smallskip
\begin{tabular}{ccccc}
  \hline\hline
  Source & $\nu_{\rm s}$ [Hz] & $\nu_{\rm K}$ [Hz] 
  & $\nu_{\rm QPO3}$ [HZ] & References \\
  \hline
  4U\,0641+91    & 323   & 900   & 22    & Stella \& Vietri \shortcite{Stella97a} \\
  \hline
  4U\,1608-52    & 233   & 800   & 20    & Yu et al. \shortcite{Yu97a} \\
  \hline
                &       & 355   & 9.0    &  \\
  4U\,1728-34    & 355   & 551   & 14.1  & Ford \& van der Klis \shortcite{Ford98b} \\
                &       & 699  & 26.5    &  \\
                &       & 1122  & 41.5    &  \\
  \hline
  KS 1731-260   & 262   & 1197  & 27    & Wijnands \& van der Klis \shortcite{Wijnands97a} \\
  \hline
  4U\,1735-44   & 326   & 1149  & 29    & Wijnands et al. \shortcite{Wijnands98b} \\
  \hline
                &       & 1050  & 5     &  \\
  Sco X-1       & 247   & 1101  & 9     & van der Klis \shortcite{VanDerKlis96a} \\
                &       & 1135  & 13    &  \\
  \hline
                &       & 644   & 5     &  \\
  GX 17+2       & 294   & 832   & 9     & Wijnands et al. \shortcite{Wijnands97b} \\
                &       & 1042  & 13    &  \\
  \hline
                &       & 731   & 5     &  \\
  Cyg X-2       & 346   & 856   & 9     & Wijnands et al. \shortcite{Wijnands98a} \\
                &       & 1007  & 13    &  \\
  \hline
                &       & 569   & 5     &  \\
  GX 340+0      & 325   & 730   & 9     & Jonker et al. \shortcite{Jonker98a} \\
                &       & 823   & 13    &  \\
  \hline
                &       & 506   & 5     &  \\
  GX 5-1        & 298   & 685   & 9     & Wijnands et al. \shortcite{Wijnands98d} \\
                &       & 889   & 13    &  \\
  \hline\hline
\end{tabular}\end{table*}
Stella \& Vietri \shortcite{Stella97a} suggested that a third oscillation
with a frequency $\nu_{\rm QPO3}$ around $\sim 10$~Hz is also produced
at the inner border of the accretion disk by Lense-Thirring
precession. The observed frequencies $\nu_{\rm QPO3}$
(s. Tab. \ref{tab:qpo}) can again be compared with theoretical neutron
star models. The neutron star models are constructed for a given spin
frequency $\nu_{\rm s}$ and a fixed, unfortunately unknown, star mass
$M$. The obtained monotone relations $\nu_{\rm K}(r,\theta=\pi/2)$ and
$\nu_{\rm LT}(r,\theta=\pi/2)$ are transformed into the relation
$\nu_{\rm LT}(\nu_{\rm K})$. Since the Lense-Thirring precession
frequency $\nu_{\rm LT}$ is, in first approximation, proportional to
$\nu_{\rm s}$ the relation $\nu_{\rm LT}/\nu_{\rm s}$ as function of
$\nu_{\rm K}$ is shown in Fig. \ref{fig:rot.qpo} for all EOSs. The
observations are shown as circles. The curves depend only weakly on
the neutron star mass. One can therefore draw conclusions from the
comparison of the theoretical curves with the observations, although
the star masses are unknown. Since both $\nu_{\rm LT}/\nu_{\rm s}$ and
$\nu_{\rm K}$ do not depend on $\nu_{\rm s}$ in first approximation,
the calculation of the theoretical curves for one specific spin
frequency, $\nu_{\rm s}=363$~Hz (this corresponds to the spin
frequency of the neutron star in 4U1728-34), is sufficient.

As can be seen in Fig. \ref{fig:rot.qpo}, the observations of
KS\,1731-260 and 4U\,1735-44 agree with the theoretical curves of
group 1, which contains the stiffest EOSs of our sample. This is in
agreement with the results of Stella \& Vietri
\shortcite{Stella97a}. However, the other observations, including the
observations of the Z-sources which are not shown, lie above all
theoretical curves. The observed frequency $\nu_{\rm QPO3}$ is thus
higher than the Lense-Thirring precession frequency even for the
stiffest EOSs. If one assume that the observed frequencies correspond
to the first overtone of the Lense-Thirring precession, the
Atoll-source observations are in the range of the theoretical curves.

In the case of the Z-sources, the detected frequencies $\nu_{\rm
QPO3}$ are not only too large in most sources, but additionally the
slope of the relation $\nu_{\rm QPO3}(\nu_{\rm QPO1})$ of the Z-source
observations is much higher than the slope of $\nu_{\rm LT}(\nu_{\rm
K})$ for the theoretically determined models. Even the assumption,
that the first overtone of the Lense-Thirring precession frequency
$\nu_{\rm LT}$ is detected \cite{Stella97a} cannot explain these
discrepancies.

\section{Conclusions} \label{sec:concl}

In this paper, we derived models of rapidly rotating neutron stars and
strange stars by solving the general relativistic structure equations
for a broad collection of modern EOSs. We compared the space time
geometry of these models with recently discovered QPOs in the X-ray
brightness of LMXBs.

If one follows the general beat-frequency interpretation of the
kilohertz-QPOs, i.e. that the higher frequency QPO originates at a
stable circular orbit, one can constrain the mass of the neutron star
to a range which depends on the EOS. This mass range is for all sources
and for all EOSs consistent with a canonical mass, $M=1.4\,M_\odot$. As
it was stated by Miller et al. \shortcite{Miller98b} and Thampan et
al. \shortcite{Thampan98a} the exact lower and upper limits of the neutron
star's mass can only be determined by using fully relativistic models
of rapidly rotating neutron stars. The exact limits differ from the
approximations with $j=0$ by roughly 10~\%.

As it was shown by Kaaret et al. \shortcite{Kaaret97a} and Zhang et
al. \shortcite{Zhang98c}, the observation of a maximum frequency $\nu_{\rm
QPO1}^{\rm max}$ of the high frequency QPO in the sources 4U\,1820-30,
4U1608-52, and 4U\,1636-536 favour the interpretation that this QPO
originates at the innermost stable orbit. If this interpretation is
correct, the mass of the neutron star can be exactly (within the
observational errors) determined for a given EOS. The approximately
obtained mass, $M=2.07\,M_\odot$, of the source 4U\,1820-30 is larger
than the maximum mass of most of the considered EOSs. This conclusion
is even strengthened if the observed maximum frequency $\nu_{\rm
QPO1}^{\rm max}$ is compared with the exact neutron star models. The
only allowed EOSs of our broad collection are then the \UVU\ and
\RHFpn, which both describe neutron star matter as consisting of
nucleons and leptons only. The derived masses of the three sources lie
in the narrow range between $M=1.92\,M_\odot$ and $2.25\,M_\odot$.

This result is also of some importance for the nature of the object
left in the supernova 1987A. During the first ten seconds after the
supernova, neutrinos were detected \cite{Chevalier97a}. This means
that a protoneutron star was formed in the supernova. The fact that up
to now no pulsar emission could be detected was interpreted by Bethe
\& Brown \shortcite{Bethe95a} that the protoneutron star collapsed to a
black hole when the star became transparent to neutrinos after roughly
10~s. The estimated value of the baryonic mass $M_{\rm B}\sim
1.63-1.76\,M_\odot$ \cite{Bethe95a,Thielemann94a} gives thus an upper
limit to the maximum gravitational mass of a neutron star: $M_{\rm
max}\lesssim 1.6\,M_\odot$. This limit is in clear contradiction to
the derived mass of, e.g. 4U\,1820-30.

It is generally believed that neutron stars are born with a mass about
$1.4-1.5\,M_\odot$. Neutron stars in LMXBs would therefore accrete
$0.4-0.8\,M_\odot$ during their lifetime. It seems reasonable to
assume that some of the neutron stars in LMXBs accrete enough matter
to reach the maximally stable mass ($M_{\rm max}=2.20\,M_\odot$ for
\UVU, $M_{\rm max}=2.44\,M_\odot$ for \RHFpn). The neutron star would
then collapse to a black hole.

The interpretation of the QPO with frequencies $\nu_{\rm QPO3}$ about
10~Hz as Lense-Thirring precession frequency of the accretion disk
seems not to be consistent with the theoretical star models, unless
one assumes that the first overtone of the precession is observed. In
the case of Z-sources however, the necessary ratio of the frame
dragging frequency and spin frequency, and thus the moment of inertia,
of the models if too small compared to the observed frequency
$\nu_{\rm QPO3}$ or half of it.

Our general conclusion is that the observations of kilohertz QPOs in
LMXBs provide us another powerful tool for probing the interior of
neutron stars. Compared to the other tools like observations of the
maximum mass \cite{Kerkwijk95a}, the limiting spin period
\cite{Friedman86a,Weber92a}, and cooling simulations
\cite{Tsuruta66,Schaab95a,Page97a}, the derived constraints,
especially in the sonic-point-interpretation, are rather strong. The
lower limit $M_{\rm max}\gtrsim 2.15\,M_\odot$ is only consistent with
two EOSs, \UVU\ and \RHFpn, which are relatively stiff at high
densities. Their stiff behaviour at high densities seems to be only
possible if the neutron star matter consists of neutrons, protons, and
leptons only. At the most, a small admixture of hyperons may be
allowed. However, one has to admit that such a composition somehow
contradicts our conception of neutron star matter, since
fieldtheoretical models lead more or less inevitably to a more
complex composition at high density. If the given interpretation is
correct, one has therefore to reinvestigate the problem of super
dense neutron star matter by dropping, for instance, the standard
assumptions about the coupling of the hyperons in such regimes.

This specific conclusion depends however on the interpretation of the
maximum frequency of the kilohertz QPO within the
sonic-point-model. As a kind of beat-frequency model, this model
predicts a constant frequency separation $\Delta\nu=\nu_{\rm
QPO1}-\nu_{\rm QPO2}$. Recently two further examples, where this
constancy is not observed, has been discovered: 4U\,1608-52 and
4U\,1735-44 \cite{Mendez98b,Ford98c}. Moreover, the frequency
separation in the Atoll source 4U\,1636-536 seems not to be consistent
with the half of the frequency of the QPO in type I bursts
\cite{Mendez98c}. Though the beat-frequency models are the most
hopeful candidates in explaining the QPO-phenomenology it has to be
awaited how the variation of the frequency separation and deviation
from the QPO frequency in bursts can be incorporated to these models.

\section*{Acknowledgements}

One of us (Ch.~S.) gratefully acknowledges the Bavarian State for
financial support. We would like to thank Norman Glendenning and
J{\"u}rgen Schaffner-Bielich for providing us tables of their EOSs.


\begin{thebibliography}{84}
\expandafter\ifx\csname natexlab\endcsname\relax\def\natexlab#1{#1}\fi

\bibitem[\protect\citename{Alcock et~al., }1986]{Alcock86}
Alcock C., Farhi E., Olinto A.V., 1986, ApJ, 310, 261

\bibitem[\protect\citename{Balberg \& Gal, }1997]{Balberg96a}
Balberg S., Gal A., 1997, Nucl. Phys. A, 625, 435

\bibitem[\protect\citename{Balberg et~al., }1998]{Balberg98a}
Balberg S., Lichtenstadt I., Cook G.B., 1998, Roles of hyperons in neutron
  stars, to be published in ApJS, preprint astro-ph/9810361

\bibitem[\protect\citename{Batty et~al., }1994]{Batty94a}
Batty C.J., Friedman E., Gal A., 1994, Phys. Lett. B, 335, 273

\bibitem[\protect\citename{Bethe \& Brown, }1995]{Bethe95a}
Bethe H.A., Brown G.E., 1995, ApJ, 445, L129

\bibitem[\protect\citename{Bodmer, }1971]{Bodmer71}
Bodmer A., 1971, Phys. Rev. D, 4, 1601

\bibitem[\protect\citename{Bonazzola et~al., }1993]{Bonazzola93a}
Bonazzola S., Gourgoulhon E., Sagado M., Marck J., 1993, A\&A, 278, 421

\bibitem[\protect\citename{B{\"o}rner, }1973]{Boerner73b}
B{\"o}rner G., 1973, in {\em Ergebnisse der exakten Naturwissenschaften, Band
  69\/}, Springer, Berlin, pp. 1--67

\bibitem[\protect\citename{Caldwell \& Friedman, }1991]{Caldwell91}
Caldwell R., Friedman J., 1991, Phys. Lett., 264B, 143

\bibitem[\protect\citename{Chevalier, }1997]{Chevalier97a}
Chevalier R.A., 1997, Sci, 276, 1374

\bibitem[\protect\citename{Cook et~al., }1992]{Cook92a}
Cook G.B., Shapiro S.L., Teukolsky S.A., 1992, ApJ, 398, 203

\bibitem[\protect\citename{Datta et~al., }1998]{Datta98a}
Datta B., Thampan A.V., Bombacci I., 1998, A\&A, 334, 943

\bibitem[\protect\citename{Farhi \& Jaffe, }1984]{Farhi84}
Farhi E., Jaffe R., 1984, Phys. Rev. D, 30, 2379

\bibitem[\protect\citename{{Ford} et~al., }1997]{Ford97a}
{Ford} E., {Kaaret} P., {Tavani} M., {Barret} D., {Bloser} P., {Grindlay} J.,
  {Harmon} B.A., {Paciesas} W.S., {Zhang} S.N., 1997, ApJ, 475, L123

\bibitem[\protect\citename{Ford \& {van der Klis}, }1998]{Ford98b}
Ford E.C., {van der Klis} M., 1998, ApJ, 506, L39

\bibitem[\protect\citename{Ford et~al., }1998{\natexlab{a}}]{Ford98a}
Ford E.C., {van der Klis} M., Kaaret P., 1998{\natexlab{a}}, ApJ, 498, L41

\bibitem[\protect\citename{Ford et~al., }1998{\natexlab{b}}]{Ford98c}
Ford E.C., {van der Klis} M., M{\'e}ndez M., Wijnands R., Kaaret P.,
  1998{\natexlab{b}}, ApJ, 508, L155

\bibitem[\protect\citename{Friedman et~al., }1986]{Friedman86a}
Friedman J.L., Ipser J.R., Parker L., 1986, ApJ, 304, 115

\bibitem[\protect\citename{Glendenning, }1989]{Glendenning89}
Glendenning N.K., 1989, Nucl. Phys. A, 493, 521

\bibitem[\protect\citename{Glendenning, }1990]{Glendenning90}
Glendenning N.K., 1990, Mod. Phys. Lett., A5, 2197

\bibitem[\protect\citename{Glendenning, }1995]{Glendenning95b}
Glendenning N.K., 1995, unpublished

\bibitem[\protect\citename{Glendenning, }1997]{Glendenning97c}
Glendenning N.K., 1997, Compact Stars: Nuclear Physics, Particle Physics and
  General Relativity, Springer, New-York

\bibitem[\protect\citename{Glendenning et~al.,
  }1995{\natexlab{a}}]{Glendenning95c}
Glendenning N.K., Kettner C., Weber F., 1995{\natexlab{a}}, Phys. Rev. C, 51,
  1790

\bibitem[\protect\citename{Glendenning et~al.,
  }1995{\natexlab{b}}]{Glendenning94a}
Glendenning N.K., Kettner C., Weber F., 1995{\natexlab{b}}, Astrophysical
  Journal, 450, 253

\bibitem[\protect\citename{Haensel \& Pichon, }1994]{Haensel94a}
Haensel P., Pichon B., 1994, A\&A, 283, 313

\bibitem[\protect\citename{Huber, }1998]{Huber98a}
Huber H., 1998, Ph.D. thesis, University of Munich, unpublished

\bibitem[\protect\citename{Huber et~al., }1998]{Huber97a}
Huber H., Weber F., Weigel M.K., Schaab C., 1998, Int. J. Mod. Phys. E, 7, 301

\bibitem[\protect\citename{Jonker et~al., }1998]{Jonker98a}
Jonker P.G., Wijnands R., {van der Klis} M., Psaltis D., Kuulkers E., Lamb
  F.K., 1998, ApJ, 499, L191

\bibitem[\protect\citename{Kaaret et~al., }1997]{Kaaret97a}
Kaaret P.K., Ford E., Chen K., 1997, ApJ, 480, L27

\bibitem[\protect\citename{Lamb et~al., }1998]{Lamb97a}
Lamb F.K., Miller M.C., Psaltis D., 1998, in {\em Accretion Processes in
  Astrophysical Systems: Some Like It Hot\/}, AIP Conference Proceedings 431

\bibitem[\protect\citename{Lense \& Thirring, }1918]{Lense18a}
Lense J., Thirring H., 1918, Phys. Z., 19, 156

\bibitem[\protect\citename{Madsen, }1998]{Madsen98a}
Madsen J., 1998, How to identify a strange star, astro-ph/9806032

\bibitem[\protect\citename{Madsen \& Haensel, }1991]{Aarhus91}
Madsen J., Haensel P., eds., 1991, Strange Quark Matter in Physics and
  Astrophysics, Proceedings of the International Workshop, Aarhus, Denmark,
  vol.~24

\bibitem[\protect\citename{Madsen \& Olesen, }1991]{Madsen91}
Madsen J., Olesen M.L., 1991, Phys. Rev. D, 43, 1069, erratum: ibid. D44 (1991)
  566

\bibitem[\protect\citename{M{\'e}ndez \& {van Paradijs}, }1998]{Mendez98c}
M{\'e}ndez M., {van Paradijs} J., 1998, Difference frequency of kilohertz qpos
  is not equal to half the burst oscillation frequency in 4u 1636-53,
  astro-ph/9808281

\bibitem[\protect\citename{M{\'e}ndez et~al., }1998{\natexlab{a}}]{Mendez97a}
M{\'e}ndez M., {van der Klis} M., {van Paradijs} J., Lewin W.H.G., Vaughen
  B.A., Kuulkers E., Zhang W., Lamb F.K., Psaltis D., 1998{\natexlab{a}}, in
  {\em Accretion Processes in Astrophysical Systems: Some Like It Hot\/}, AIP
  Conference Proceedings 431

\bibitem[\protect\citename{M{\'e}ndez et~al., }1998{\natexlab{b}}]{Mendez98a}
M{\'e}ndez M., {van der Klis} M., {van Paradijs} J., Lewin W.H.G., Vaughen
  B.A., Kuulkers E., Zhang W., Lamb F.K., Psaltis D., 1998{\natexlab{b}}, ApJ,
  494, L65

\bibitem[\protect\citename{M{\'e}ndez et~al., }1998{\natexlab{c}}]{Mendez98b}
M{\'e}ndez M., {van der Klis} M., Wijnands R., Ford E.C., {van Paradijs} J.,
  Vaughen B.A., 1998{\natexlab{c}}, ApJ, 505, L23

\bibitem[\protect\citename{Miller, }1998]{Miller98c}
Miller M.C., 1998, Evidence for antipodal hot spots during x-ray bursts from
  {4U 1636-536}, astro-ph/9809235

\bibitem[\protect\citename{Miller et~al., }1998{\natexlab{a}}]{Miller98b}
Miller M.C., Lamb F.K., Cook G.B., 1998{\natexlab{a}}, ApJ, 509, 793, subm. to
  ApJ, preprint astro-ph/9805007

\bibitem[\protect\citename{Miller et~al., }1998{\natexlab{b}}]{Miller98a}
Miller M.C., Lamb F.K., Psaltis D., 1998{\natexlab{b}}, in Scarsi L., Bradt H.,
  Giommi P., Fiore F., eds., {\em The Active X-Ray Sky: Results from BeppoSAX
  and RXTE\/}, Elsevier (Amsterdam), Nucl. Phys. B 69

\bibitem[\protect\citename{Miller et~al., }1998{\natexlab{c}}]{Miller96a}
Miller M.C., Lamb F.K., Psaltis D., 1998{\natexlab{c}}, ApJ, 508, 791

\bibitem[\protect\citename{Misner et~al., }1973]{Misner73}
Misner C.W., Thorne K.S., Wheeler J.A., 1973, Gravitation, W.H. Freeman and
  Company, New York

\bibitem[\protect\citename{Myers \& Swiatecki, }1996]{Myers96a}
Myers W.D., Swiatecki W.J., 1996, Nucl. Phys. A, 601, 141

\bibitem[\protect\citename{Negele \& Vautherin, }1973]{Negele73}
Negele J., Vautherin D., 1973, Nucl. Phys. A, 207, 298

\bibitem[\protect\citename{Page, }1998]{Page97a}
Page D., 1998, in Buccheri R., {van Paradijs} J., Alpar A., eds., {\em The Many
  Faces of Neutron Stars\/}, Kluwer (Dordrecht)

\bibitem[\protect\citename{Pethick et~al., }1995]{Pethick95}
Pethick C., Ravenhall P., Lorenz C., 1995, Nucl. Phys. A, 584, 675

\bibitem[\protect\citename{Psaltis et~al., }1998]{Psaltis98a}
Psaltis D., Mendez M., Wijnands R., Homan J., Jonker P.G., {van der Klis} M.,
  Lamb F.K., Kuulkers E., {van Paradijs} J., Lewin W.H.G., 1998, ApJ, 501, L95

\bibitem[\protect\citename{Schaab, }1999]{Schaab97a}
Schaab C., 1999, Struktur und Thermische Entwicklung von Neutronensternen und
  Strange Sternen, GCA-Verlag (Herdecke), ISBN 3-928973-53-3, PhD-Thesis

\bibitem[\protect\citename{Schaab et~al., }1996]{Schaab95a}
Schaab C., Weber F., Weigel M.K., Glendenning N.K., 1996, Nucl. Phys. A, 605,
  531

\bibitem[\protect\citename{Schaffner \& Mishustin, }1996]{Schaffner95a}
Schaffner J., Mishustin I.N., 1996, Phys. Rev. C, 53, 1416

\bibitem[\protect\citename{Smarr \& York, }1978]{Smarr78a}
Smarr L., York J.W., 1978, Phys. Rev. D, 17, 2529

\bibitem[\protect\citename{Smith et~al., }1997]{Smith97a}
Smith D.A., Morgan E.H., Bradt H., 1997, ApJ, 479, L137

\bibitem[\protect\citename{Stella \& Vietri, }1998]{Stella97a}
Stella L., Vietri M., 1998, ApJ, 492, L59

\bibitem[\protect\citename{Strobel et~al., }1997]{Strobel96a}
Strobel K., Weber F., Schaab C., Weigel M.K., 1997, Int. J. Mod. Phys. E, 6,
  669

\bibitem[\protect\citename{Strohmayer, }1998]{Strohmayer98a}
Strohmayer T.E., 1998, in {\em Accretion Processes in Astrophysical Systems:
  Some Like It Hot\/}, AIP Conference Proceedings 431

\bibitem[\protect\citename{Strohmayer et~al., }1996]{Strohmayer96a}
Strohmayer T.E., Zhang W., Swank J.H., Smale A., Titarchuk L., Day C., 1996,
  ApJ, 469, L9

\bibitem[\protect\citename{Strohmayer et~al., }1998]{Strohmayer98b}
Strohmayer T.E., Zhang W., Swank J.H., White N.E., 1998, ApJ, 498, L135

\bibitem[\protect\citename{Thampan et~al., }1999]{Thampan98a}
Thampan A.V., Bhattacharya D., Datta B., 1999, MNRAS, 302, L69

\bibitem[\protect\citename{Thielemann et~al., }1996]{Thielemann94a}
Thielemann F.K., Nomoto K., Hashimoto M., 1996, ApJ, 460, 408

\bibitem[\protect\citename{Tsuruta, }1966]{Tsuruta66}
Tsuruta S., 1966, Canadian Journal of Physics, 44, 1863

\bibitem[\protect\citename{{van der Klis}, }1998]{VanDerKlis97a}
{van der Klis} M., 1998, in Buccheri R., {van Paradijs} J., Alpar A., eds.,
  {\em The Many Faces of Neutron Stars\/}, Kluwer (Dordrecht)

\bibitem[\protect\citename{{van der Klis} \& Wijnands, }1997]{VanDerKlis97b}
{van der Klis} M., Wijnands R.A.D., 1997, ApJ, 481, L97

\bibitem[\protect\citename{{van der Klis} et~al., }1996]{VanDerKlis96a}
{van der Klis} M., Swank J.H., Zhang W., Jahoda K., Morgan E.H., Lewin W.H.G.,
  Vaughan B., {van Paradijs} J., 1996, ApJ, 469, L1

\bibitem[\protect\citename{{van Kerkwijk} et~al., }1995]{Kerkwijk95a}
{van Kerkwijk} M.H., {van Paradijs} J., Zuiderwijk E.J., 1995, A\&A, 303, 497

\bibitem[\protect\citename{Weber \& Glendenning, }1992]{Weber92a}
Weber F., Glendenning N.K., 1992, ApJ, 390, 541

\bibitem[\protect\citename{Weber \& Glendenning, }1993]{Weber91}
Weber F., Glendenning N.K., 1993, in Feng D.H., He G.Z., Li X.Q., eds., {\em
  Proceedings of the Nankai Summer School, ``Astrophysics and Neutrino
  Physics'', Tanjin, China, June 17-27, 1991\/}, World Scientific, Singapore,
  pp. 64--183

\bibitem[\protect\citename{Weber \& Weigel, }1989]{Weber89}
Weber F., Weigel M., 1989, Nucl. Phys. A, 505, 779

\bibitem[\protect\citename{Weber et~al., }1995]{Weber94}
Weber F., Kettner C., Weigel M.K., Glendenning N.K., 1995, in Vassiliadis G.,
  Panagiotou A., Kumar S., Madsen J., eds., {\em Proceedings of the
  International Symposium on Strangeness and Quark Matter\/}, World Scientific,
  pp. 308

\bibitem[\protect\citename{Weber et~al., }1997{\natexlab{a}}]{Weber96b}
Weber F., Schaab C., Weigel M.K., Glendenning N.K., 1997{\natexlab{a}}, in
  Giovannelli F., Mannocci G., eds., {\em Vulcano Workshop 1996 Frontier
  Objects in Astrophysics and Particle Physics, May 27 -- June 1, Vulcano\/},
  Italian Physical Society, Bologna (Italia), p.~87

\bibitem[\protect\citename{Weber et~al., }1997{\natexlab{b}}]{Weber96a}
Weber F., Schaab C., Weigel M.K., Glendenning N.K., 1997{\natexlab{b}}, in
  St\"ocker H., Gallmann A., Hamilton J.H., eds., {\em Procceedings of the
  International Conference on Nuclear Physics at the Turn of the Millenium:
  Structure of Vacuum and Elementary Matter, March 10-16, 1996, Wilderness,
  South Africa\/}, World Scientific, Singapore, p. 322

\bibitem[\protect\citename{Wijnands \& {van der Klis}, }1997]{Wijnands97a}
Wijnands R.A.D., {van der Klis} M., 1997, ApJ, 482, L65

\bibitem[\protect\citename{Wijnands et~al., }1997]{Wijnands97b}
Wijnands R.A.D., Homan J., {van der Klis} M., M\'endez M., Kuulkers E., {van
  Paradijs} J., Lewin W.H.G., Lamb F.K., Psaltis D., Vaughan B., 1997, ApJ,
  490, L157

\bibitem[\protect\citename{Wijnands et~al., }1998{\natexlab{a}}]{Wijnands98a}
Wijnands R.A.D., Homan J., {van der Klis} M., Kuulkers E., {van Paradijs} J.,
  Lewin W.H.G., Lamb F.K., Psaltis D., Vaughan B., 1998{\natexlab{a}}, ApJ,
  493, L87

\bibitem[\protect\citename{Wijnands et~al., }1998{\natexlab{b}}]{Wijnands98d}
Wijnands R.A.D., M{\'e}ndez M., {van der Klis} M., Psaltis D., Kuulkers E.,
  Lamb F.K., 1998{\natexlab{b}}, ApJ, 504, L35

\bibitem[\protect\citename{Wijnands et~al., }1998{\natexlab{c}}]{Wijnands98b}
Wijnands R.A.D., {van der Klis} M., M{\'e}ndez M., {van Paradijs} J., Lewin
  W.H.G., Lamb F.K., Vaughan B., Kuulkers E., 1998{\natexlab{c}}, ApJ, 495, L39

\bibitem[\protect\citename{Wiringa et~al., }1988]{Wiringa88}
Wiringa R., Fiks V., Fabrocini A., 1988, Phys. Rev. C, 38, 1010

\bibitem[\protect\citename{Witten, }1984]{Witten84}
Witten E., 1984, Phys. Rev. D, 30, 272

\bibitem[\protect\citename{Yu et~al., }1997{\natexlab{a}}]{Yu97a}
Yu W., Zhang S.N., Harmon B.A., Paciesas W.S., Robinson C.R., Grindlay J.E.,
  Bloser P., Barret D., Ford E.C., Tavani M., Kaaret P., 1997{\natexlab{a}}, in
  Dermer C.D., Strickman M.S., Kurfess J.D., eds., {\em Proceedings of the
  Fourth Compton Symposium\/}, AIP Conference Proceedings 410, p. 734

\bibitem[\protect\citename{Yu et~al., }1997{\natexlab{b}}]{Yu97b}
Yu W., Zhang S.N., Harmon B.A., Paciesas W.S., Robinson C.R., Grindlay J.E.,
  Bloser P., Barret D., Ford E.C., Tavani M., Kaaret P., 1997{\natexlab{b}},
  ApJ, 490, L153

\bibitem[\protect\citename{Zhang et~al., }1997]{Zhang97a}
Zhang W., Strohmayer T.E., Swank J.H., 1997, ApJ, 482, L167

\bibitem[\protect\citename{Zhang et~al., }1998{\natexlab{a}}]{Zhang98a}
Zhang W., Jahoda K., Kelley R.L., Strohmayer T.E., Swank J.H., Zhang S.N.,
  1998{\natexlab{a}}, ApJ, 495, L9

\bibitem[\protect\citename{Zhang et~al., }1998{\natexlab{b}}]{Zhang98c}
Zhang W., Smale A.P., Strohmayer T.E., Swank J.H., 1998{\natexlab{b}}, ApJ,
  500, L171

\bibitem[\protect\citename{Zhang et~al., }1998{\natexlab{c}}]{Zhang98b}
Zhang W., Strohmayer T.E., Swank J.H., 1998{\natexlab{c}}, ApJ, 500, L167

\end{thebibliography}

\end{document}